\documentclass[useAMS,usenatbib,referee]{biom}

\usepackage{bm}
\usepackage{indentfirst}
\usepackage{amsmath}
\usepackage{graphicx}
\usepackage{subfigure}
\usepackage{threeparttable}
\usepackage{booktabs}
\usepackage{multirow, color, dsfont}
\usepackage{mathrsfs}
\usepackage{setspace}
\usepackage{float}
\usepackage{bbm}
\usepackage{url}

 \newcommand{\bitem}{\begin{itemize}}
 \newcommand{\eitem}{\end{itemize}}

\usepackage[figuresright]{rotating}

\title[Sloan et al.]{A Repeated Measures Approach to Pooled and Calibrated Biomarker Data}

\author
{Abigail Sloan$^1$, Chao Cheng$^2$, Bernard Rosner$^1$, Regina G. Ziegler$^3$, Stephanie A. \\ \textbf{Smith-Warner$^4$, Molin Wang$^{1,5,6}$}\emailx{abbysloan@gmail.com} \\
Departments of Biostatistics$^1$, Nutrition$^4$, Epidemiology$^5$,  Harvard T. H. Chan School of Public \\ Health, 677 Huntington Ave, Boston, MA, 02215 \\
$^2$Center for Methods in Implementation and Prevention Science, Yale School of  Public \\ Health, New Haven, CT \\
$^3$Division of Cancer Epidemiology and Genetics, National Cancer Institute, National Institutes \\ of Health, Bethesda, MD \\
$^6$Channing Division of Network Medicine, Department of Medicine, Brigham and Women's Hospital, \\ and Harvard Medical School, Boston, MA
  }

\begin{document}

%  This will produce the submission and review information that appears
%  right after the reference section.  Of course, it will be unknown when
%  you submit your paper, so you can either leave this out or put in 
%  sample dates (these will have no effect on the fate of your paper in the
%  review process!)

\date{{\it Received XX} 20XX. {\it Revised XX} 20XX.  {\it
Accepted XX} 20XX.}

\pagerange{\pageref{firstpage}--\pageref{lastpage}} 
\volume{63}
\pubyear{2021}
\artmonth{February}

\doi{xx.xxxx/j.xxxx-xxxx.xxxx.xxxxx.x}

\label{firstpage}

%  put the summary for your paper here

\begin{abstract}
Participant level meta-analysis across multiple studies increases the sample size for pooled analyses, thereby improving precision in effect estimates and enabling subgroup analyses. For analyses involving biomarker measurements as an exposure of interest, investigators must first calibrate the data to address measurement variability arising from usage of different laboratories and/or assays. In practice, the calibration process involves reassaying a random subset of biospecimens from each study at a central laboratory and fitting models that relate the study-specific ``local" and central laboratory measurements. Previous work in this area treats the calibration process from the perspective of measurement error techniques and imputes the estimated central laboratory value among individuals with only a local laboratory measurement. In this work, we propose a repeated measures method to calibrate biomarker measurements pooled from multiple studies with study-specific calibration subsets. We account for correlation between measurements made on the same person and between measurements made at the same laboratory. We demonstrate that the repeated measures approach provides valid inference, and compare it to existing calibration approaches grounded in measurement error techniques in an example describing the association between circulating vitamin D and stroke.
\end{abstract}


\maketitle

\section{Introduction \label{section intro}}

Participant level meta-analysis remains an important tool for increasing sample size in pooled analyses [\cite{Debray13,Key10,Smith06, Tworoger06}]. Individual studies may have insufficient data to estimate exposure-disease risk in particular subgroups or inadequate power to detect statistically significant associations. Furthermore, participant level meta-analysis allows maximal usage of the existing evidence across multiple studies, resulting in more precise effect estimates over a broader range of exposures. The increasing number of epidemiologic data consortia over time attests to the advantages offered by pooled individual level meta-analysis [\cite{McCullough18,Crowe14, Key15}].

Pooled analyses involving biomarker data can be challenging due to potential assay or laboratory variability across studies. If investigators do not harmonize data prior to completing an individual level meta-analysis, effect estimates are likely to be biased toward the null [\cite{Tworoger06}]. For example, blood circulating 25-hydroxyvitamin D measurements (25(OH)D) can vary as much as 40\% across assays [\cite{Barake12}]. Thus, investigators must address inter-study and inter-assay variability in their statistical analyses in order to complete pooled analyses of biomarker data.  

Investigators typically account for variable biomarker measurements by obtaining calibration data and implementing calibration models in either a one-stage or two-stage approach  [\cite{Sloan18, Sloan19, Gail16, Debray13,Freedman11}]. Calibration data consists of re-assayed biomarker measurements for a random subset of individuals in each study at a central laboratory. Under the one-stage approach, the central laboratory calibration measurements are used to harmonize all biomarker measurements prior to completing a single statistical analysis of data pooled across all studies (i.e. individual level meta-analysis) [\cite{Sloan18, Sloan19}]. These authors described this strategy as the `aggregated approach' and derived two methods, the internalized method and the full calibration method. Under the two-stage approach, study-specific effect estimates are obtained and calibrated using regression calibration in the first stage, followed by meta-analysis of the results in the second stage [\cite{Sloan18, Sloan19}]. 

Although the one and two-stage calibration approaches previously described by Sloan et al. (2018, 2019) provide valid inference for the biomarker-disease association under appropriate assumptions, they still face some critical limitations. First, these methods regard the biomarker measurements obtained by the central laboratory as a gold standard. In practice, central laboratories are often chosen for logistical reasons and may not offer more accurate or less variable measurements than the local laboratories. The central laboratory measurements simply serve as a benchmark for all measurements. Second, these methods assume a specific form of linear relationship between the central and local laboratory measurements. Other approaches, like that proposed by Freedman et al. (2011), do not propagate uncertainty from the calibration process into the estimated variability for the effect estimate.

In response to these limitations, we develop a repeated measures approach to pooled and calibrated biomarker data. Instead of using measurement-error models to estimate missing central laboratory data like previous techniques, we estimate a true biomarker value while accounting for correlation within measurements made at the same laboratory or on the same person. 

In this paper, we develop a repeated measures approach for calibrating biomarker data pooled from multiple studies to analyze the association between a continuous biomarker measurement and binary disease outcome. Section 2 presents the models and methods, including a logistic biomarker-disease model and a linear mixed model for computing calibrated biomarker measurements. Simulation studies in Section 3 consider the operating characteristics of the repeated measures approach under various conditions. Section 4 implements the method in an example relating 25(OH)D to stroke incidence in the Health Professionals Follow-up Study (HPFS), Nurses' Health Study I (NHS1), and Nurses' Health Study II (NHS2), and compares the result to the results obtained by following the one or two-stage approach of Sloan et al. (2018). Section 5 discusses our conclusions.

\section{Methods}

\subsection{Notation and models}
\label{secnotation}

For individual $k$ of study $j$, let $X_{jk}$ and $Y_{jk}$ denote the true biomarker value and binary disease outcome respectively. We model the biomarker-disease association in the pooled analysis using a logistic regression model such that

\begin{align}
\label{logit}
    \textrm{logit}(P(Y_{jk}=1|X_{jk},\bm{Z}_{jk})) = \beta_{0j} + \beta_x X_{jk} + \bm{\beta}_z^T \bm{Z}_{jk},
\end{align}

\noindent where $\beta_{0j}$ is a study-specific intercept, $\bm{Z}_{jk}$ contains additional covariates, and superscript $T$ denotes transpose. Without further specification, all vectors are column vectors. We assume that at least two studies contribute to the analysis that use different laboratories. Our goal is to perform inference on $\beta_x$, the log odds ratio (OR) describing the relationship between the biomarker and outcome controlling for the covariates.

Let $\bm{W}_{jk}$ be a vector of $p$ variables that are associated with $X_{jk}$, which may also be a subset of $\bm{Z}_{jk}$. $\bm{W}_{jk}$ may also include additional variables that are informative to $X_{jk}$ but are independent of $Y_{jk}$ conditional on $X_{jk}$, $\bm{Z}_{jk}$, and any measurements of the biomarker, namely $\bm{H}_{jk}$ (see Equation 3). Suppose $m$ studies contribute to the analysis such that $j=1,\dots,m$ and $m \geq 2$. We model the true laboratory measurement as a function of covariates such that

\begin{equation}
\begin{split}
\label{defineX}
X_{jk} & = \alpha_{0j} + \bm{\tau}^T \bm{W}_{jk} + \epsilon_{xjk}, 
\end{split}
\end{equation}

\noindent where $\alpha_{0j}$ is a study-specific intercept and $\epsilon_{xjk}$ is an error term with distribution $N(0,\sigma^2_x)$. Matrix $\bm{W}_{jk}$ can contain splines or other non-linear terms for continuous variables that are associated with $X_{jk}$. If no variables are associated with $X_{jk}$, then $\bm{\tau}^T \bm{W}_{jk}$ is omitted from (\ref{defineX}) and $\alpha_{0j}$ is study-specific. It is also possible that  $\alpha_{0j}$ is not a function of study, in which case the intercept would be the same across studies. 

Each study has its own distinct local laboratory and measurements from a secondary (central) laboratory are available for a subset in each study. Each individual in the analysis has a local laboratory measurement and may have an additional central laboratory measurement if selected into the calibration subset. Let $d=0$ refer to a central laboratory and $d=j$ index the local laboratory associated with study $j$ such that $d=0,1,\dots,m$. We assume a laboratory-specific random effect $\xi_d \sim N(0,\sigma^2_{\xi})$, and conditional on the true biomarker value and this random effect, the mean of the laboratory measurement is $X_{jk}+\xi_d$. Furthermore, each measurement $H_{djk}$ can be written as

\begin{equation}
\begin{split}
\label{defineH}
H_{djk} & = X_{jk} + \xi_d + \epsilon_{djk}, 
\end{split}
\end{equation}

\noindent where $\epsilon_{djk}$ is additional measurement error with distribution $N(0,\sigma^2_d)$, and $\sigma^2_d$ is a laboratory-specific variance, $d=0,1,\dots,m$, with $j$ indexing the specific study. Note that $\xi_d$ represents a between-laboratory difference and $\epsilon_{djk}$ describes within-laboratory sampling error. Previous work of Gail et al. (2016) suggests these mean zero assumptions are reasonable. Let $\bm{H}_{jk}$ denote all measurements of $X_{jk}$ such that $\bm{H}_{jk}=[H_{0jk},H_{djk}]^T$ if the individual has central and local laboratory measurements and $\bm{H}_{jk}=H_{djk}$ if the individual has only a local laboratory measurement. Note that it is possible for multiple studies to use the same local laboratory. The method does not require $E(\xi_d)=0$ since any nonzero mean would be absorbed by the intercept in model (2) as demonstrated in Appendix G of the Supporting information. However, for convention, we assume mean zero.

\subsection{Approximate likelihood}

We derive a likelihood-based method to estimate $\beta_x$. The likelihood for model (\ref{logit}) is 

\begin{equation*}
\begin{split}
L & =  \prod_j \prod_k \frac{ \exp \{ \mathds{I}(Y_{jk}=1)(\beta_{0j} + \beta_x X_{jk}+ \bm{\beta}_z^T \bm{Z}_{jk}) \} }{  \exp\{ \beta_{0j} + \beta_x X_{jk}+ \bm{\beta}_z^T \bm{Z}_{jk} \}+1}.   
\end{split}
\end{equation*}

\noindent Because we cannot measure $X_{jk}$ for all subjects, we develop an approximate likelihood that uses an estimate of $X_{jk}$ based on the available laboratory measurements and covariates. Denote this surrogate measure as $\tilde{X}_{jk}=E[X_{jk}|\bm{H}_{jk},\bm{W}_{jk}]$.

We leverage a surrogacy assumption to estimate an approximate likelihood, which states $P(Y_{jk}|X_{jk},\bm{H}_{jk},  \bm{Z}_{jk})=P(Y_{jk}|X_{jk},\bm{Z}_{jk})$. In effect, this statement asserts that the laboratory measurements $\bm{H}_{jk}$ contain no additional information about $Y_{jk}$ if the true measurement $X_{jk}$ is already known.

As described previously, we can perform a second order Taylor series approximation in the exact likelihood with respect to $X_{jk}$ about $E[X_{jk}|\bm{H}_{jk},\bm{W}_{jk}]$ to derive an approximate likelihood [\cite{Carroll06, Rosner89}]. We derive this likelihood in Section A of the Supporting information. As long as $|\beta_x|$ and $Var(X_{jk}|\bm{H}_{jk},\bm{W}_{jk})$ are not too large (i.e. the effect of the biomarker is not very large and the laboratory measurements are not too noisy), the approximation will hold. Under these conditions, we can fit the following approximate model to obtain $\beta_x$ estimates:

\begin{align}
\label{tildelogit}
    \textrm{logit}(P(Y_{jk}=1|\tilde{X}_{jk},\bm{Z}_{jk}) = \beta_{0j} + \beta_x \tilde{X}_{jk} + \bm{\beta}_z^T \bm{Z}_{jk}.
\end{align}

\noindent The presence of $\tilde{X}_{jk}$ in the approximate model motivates the development of the calibration models $E[X_{jk}|\bm{H}_{jk},\bm{W}_{jk}]$.

\subsection{Calibration step}
\label{seccalstep}

Based on our specification, the conditional distribution of $X_{jk}$ given the covariates $\bm{W}_{jk}$ and available laboratory measurements $\bm{H}_{jk}$ follows a normal distribution. For individuals in the calibration subset who have both local and central laboratory measurements, we have

\begin{align*}
{\scriptstyle
X_{jk} | \bm{H}_{jk}, \bm{W}_{jk} } & {\scriptstyle \sim N \bigg( w_j \left[ \frac{\sigma^2_0}{\sigma^2_0 + \sigma^2_j}(H_{djk}-\xi_j) + \frac{\sigma^2_j}{\sigma^2_0+\sigma^2_j} (H_{0jk}-\xi_0)  \right] + (1-w_j)(\alpha_{0j}+\bm{\tau}^T \bm{W}_{jk}) , 
 \sigma^2_x(1-w_j) \bigg),} 
\end{align*}

\noindent where $w_j=\sigma^2_x / (\sigma^2_x + \frac{\sigma^2_j \sigma^2_0}{\sigma^2_j+\sigma^2_0})$. For individuals who have only a local laboratory measurement, we have

\begin{equation*}
\begin{split}
X_{jk} | \bm{H}_{jk}, \bm{W}_{jk} & \sim N\left( \frac{\sigma^2_x}{\sigma^2_x+\sigma_j^2}(H_{djk}-\xi_j) + \frac{\sigma^2_j}{\sigma^2_x+\sigma_j^2} (\alpha_{0j}+\bm{\tau}^T\bm{W}_{jk}), \frac{\sigma^2_x \sigma^2_j}{\sigma^2_x+\sigma^2_j} \right). 
\end{split}
\end{equation*}

\noindent The means of these distributions, namely $\tilde{X}_{jk}=E[X_{jk}|\bm{H}_{jk},\bm{W}_{jk}]$, thus give formulas for the calibrated biomarker measurement. Derivations of these distributions are included in Section B of the Supporting information. We now discuss the linear mixed model (LMM) used to find the necessary estimates of the fixed effects ($\alpha_{01},\dots,\alpha_{0m},\bm{\tau}$), random effects ($\xi_0,\dots,\xi_m$), and variance terms ($\sigma^2_0,\dots,\sigma^2_m,\sigma^2_{\xi},\sigma^2_{x}$) in order to estimate $\tilde{X}_{jk}$ for use in model (\ref{tildelogit}).

\subsection{Linear mixed effects model and estimation}
\label{seclmmdef}

Based on the specification in models (\ref{defineX}) and (\ref{defineH}), we can rewrite each measurement as $H_{djk} = (\alpha_{0j} + \bm{\tau}^T \bm{W}_{jk}) + \xi_d +  \delta_{djk}$, where $\delta_{djk}=\epsilon_{xjk}+\epsilon_{djk}$. Thus, each measurement is a sum of fixed effects $(\alpha_{0j} + \bm{\tau}^T \bm{W}_{jk})$, a lab-specific random effect $\xi_d$, and an error term $\delta_{djk}$. Assume $N$ total measurements from $n$ total individuals will be included in the final analysis.

Define a vector of fixed effects parameters $\bm{\theta}=[\alpha_{01},\alpha_{02},\dots,\alpha_{0m},\bm{\tau}^T]^T$, a vector of random effects parameters $\bm{\xi}=[\xi_0, \xi_1, \dots, \xi_m]^T$, and a vector of variance terms $\bm{\sigma^2}=[\sigma^2_0,\sigma^2_1,\dots,\sigma^2_m, \\ \sigma^2_{\xi},\sigma^2_{x}]^T$. Let $\bm{C}_{djk}$ correspond to the vector of covariates in model (\ref{defineX}), including indicators for the study-specific intercept terms and $\bm{W}_{jk}$. Let $\bm{U}_{djk}$ be a vector of length $m+1$ indicating laboratory cluster membership for measurement $H_{djk}$ (full details in Section C of the Supporting information).

Let measurement vector $\bm{H}$ and error vector $\bm{\delta}$ contain all $N$ measurements and error terms respectively. The LMM is $H_{djk} = \bm{C}_{djk}^T \bm{\theta} + \bm{U}_{djk}^T \bm{\xi} + \delta_{djk}$ in scalar notation and
$\bm{H} = \bm{C} \bm{\theta} + \bm{U} \bm{\xi} + \bm{\delta}$ in matrix notation. We assume that the random effects are independent of the error terms, and that the random effects are independent of each other such that the joint distribution is

\begin{align*}
\begin{pmatrix} \bm{\xi} \\ \bm{\delta} \end{pmatrix} \sim N\left( \begin{pmatrix} \bm{0} \\ \bm{0} \end{pmatrix}, 
\begin{pmatrix} 
\bm{G} & \bm{0} \\
\bm{0} & \bm{R} \\
\end{pmatrix} \right),
\end{align*}

\noindent where $\bm{G}$ is an $m+1$ square diagonal matrix with $\sigma^2_{\xi}$ on the diagonal. Matrix $\bm{R}$ is a nearly diagonal matrix of dimension $N \times N$. The elements of the diagonal entries of $\bm{R}$ are $Var(\delta_{djk})=\sigma^2_x + \sigma^2_d$, $d=0,1,\dots,m$. The nonzero off-diagonal terms of $\bm{R}$ describe the covariance between repeated measurements on the same person, namely $Cov(\delta_{0jk}, \delta_{jjk})=\sigma^2_x$. 

We provide one final formulation of the LMM to facilitate discussion of the estimation process in Section \ref{secestimate3}. Suppose we rewrite $\bm{H}$ as $\bm{H}=\bm{C \theta} + \bm{\delta}^{*}$, where $\bm{\delta}^{*}=\bm{U \xi} + \bm{\delta}$. Under this specification, $\bm{\delta}^{*} \sim N(\bm{0},\bm{V})$, where $\bm{V}=\bm{UGU}^T+\bm{R}$.

\subsection{Estimation of $(\hat{\bm{\theta}},\hat{\bm{\xi}},\hat{\bm{\sigma}}^2)$ and $\hat{\bm{\beta}}$}
\label{secestimate3}

We obtain $(\hat{\bm{\theta}},\hat{\bm{\xi}},\hat{\bm{\sigma}}^2)$ through a three step sequential estimation procedure. First, we estimate $\hat{\bm{\sigma}}^2$ through the minimum norm quadratic unbiased estimation (MINQUE) procedure of Rao (1972), which does not require estimates of the fixed effects. Second, given $\hat{\bm{\sigma}}^2$, we estimate $\hat{\bm{\theta}}$ using the fixed effects MLE formula under LMMs. Third, we plug $\hat{\bm{\theta}}$ and $\hat{\bm{\sigma}}^2$ into the empirical best linear unbiased predictor (BLUP) formula to obtain $\hat{\bm{\xi}}$. 

 For known $\bm{V}$, the fixed effects MLE in the LMM is $\tilde{\bm{\theta}} = \bm{(\bm{C}^T \bm{V}^{-1} \bm{C})^{-1} \bm{C}^T \bm{V}^{-1}} \bm{H}$. We evaluate $\bm{V}$ at $\bm{\sigma}^2=\hat{\bm{\sigma}}^2$ to provide a consistent estimate of $\bm{V}$, namely $\hat{\bm{V}}$, in order to compute $\hat{\bm{\theta}}$. The final result is $\hat{\bm{\theta}} = \bm{(\bm{C}^T \hat{\bm{V}}^{-1} \bm{C})^{-1} \bm{C}^T \hat{\bm{V}}^{-1}} \bm{H}$, where $\hat{\bm{\theta}}$ is the best linear unbiased estimator of the fixed effects $\bm{\theta}$ in the LMM for a given $\hat{\bm{V}}$ [\cite{Henderson75}]. 

In the final step, we estimate the random effects $\hat{\bm{\xi}}$ using the previously estimated $(\hat{\bm{\sigma}}^2, \hat{\bm{\theta}})$. From Henderson (1975), the BLUP of the random effects is $\tilde{\bm{\xi}} = \bm{G U}^T \bm{V}^{-1}(\bm{H} - \bm{C \theta})$.
Like before, we replace unknown quantities with their empirical versions. The resulting empirical BLUP is $\hat{\bm{\xi}} = \hat{\bm{G}} \bm{U}^T \hat{\bm{V}}^{-1}(\bm{H} - \bm{C} \hat{\bm{\theta}})$. See Sections D and E of the Supporting information for an illustrative example of the notation.

After completing this three-step estimation procedure, we use $(\hat{\bm{\theta}},\hat{\bm{\xi}},\hat{\bm{\sigma}}^2)$ to estimate the calibrated measurements $\tilde{X}_{jk}$ (Section \ref{seccalstep}). We then fit model (\ref{tildelogit}) to estimate $\hat{\beta}_x$, the log OR describing the association between the biomarker exposure measurements and binary disease outcome. Because we used $\tilde{X}_{jk}$ in place of the true $X_{jk}$ measurements, the $\widehat{Var}(\hat{\bm{\beta}})$ estimates resulting from fitting the model may be too small. We next propose a variance estimator that accounts for the extra variability introduced by calibrating the biomarker measurements.

\subsection{Variance estimation of $\hat{\bm{\beta}}$}

We use a resampling method related to the approach for multiple imputation to obtain estimates of $\widehat{Var}(\hat{\bm{\beta}})$ [\cite{Rubin04}]. We generate the $i^{th}$ pseudo dataset as follows:

\begin{enumerate}
\item Generate new fixed and random effects via $\tilde{\bm{\theta}}^{(i)}  \sim N(\hat{\bm{\theta}}, \widehat{Var}(\hat{\bm{\theta}}))$ and 
$\tilde{\bm{\xi}}^{(i)}   \sim N(\hat{\bm{\xi}}, \widehat{Var}(\hat{\bm{\xi}}))$, where $\widehat{Var}(\hat{\bm{\theta}})$ and $\widehat{Var}(\hat{\bm{\xi}})$ are the estimated variances assuming the second-moment parameters in $\bm{V}$ are known. Their formulas are given in Section F of the Supporting information.

\item Using $\tilde{\bm{\theta}}^{(i)}$ and $\tilde{\bm{\xi}}^{(i)}$, compute new $\tilde{X}_{jk}^{(i)}$ such that

\begin{align*}
\textrm{Calibration subjects: } \tilde{X}_{jk}^{(i)} & = \hat{w}_j \left[ \frac{\hat{\sigma}^2_0}{\hat{\sigma}^2_0 +\hat{\sigma}^2_j}(H_{djk}-\tilde{\xi}_j^{(i)}) 
+ \frac{\hat{\sigma}^2_j}{\hat{\sigma}^2_0 +\hat{\sigma}^2_j} (H_{0jk}-\tilde{\xi}_0^{(i)})  \right] \\
& + (1-\hat{w}_j)(\tilde{\alpha}_{0j}^{(i)}+\tilde{\bm{\tau}}^{(i)T}\bm{W}_{jk}) \\
\textrm{Noncalibration subjects: } \tilde{X}_{jk}^{(i)} & = \frac{\hat{\sigma}^2_x}{\hat{\sigma}^2_x+\hat{\sigma}_j^2}(H_{djk}-\tilde{\xi}_j^{(i)}) + \frac{\hat{\sigma}^2_j}{\hat{\sigma}^2_x+\hat{\sigma}_j^2} (\tilde{\alpha}_{0j}^{(i)}+\tilde{\bm{\tau}}^{(i)T}\bm{W}_{jk}) \\
\end{align*}

\noindent where $\hat{w}_j=\hat{\sigma}^2_x / (\hat{\sigma}^2_x + \frac{\hat{\sigma}^2_j\hat{\sigma}^2_0}{\hat{\sigma}^2_j+\hat{\sigma}^2_0})$.

\item Fit the following logistic regression model to obtain $\hat{\bm{\beta}}^{(i)}$ and $\widehat{Var}(\hat{\bm{\beta}}^{(i)})$.

\begin{align*}
\textrm{logit}(P(H_{djk}=1|\tilde{X}_{jk}^{(i)},\bm{Z}_{jk})) = \beta_{0j}^{(i)} + \beta_x^{(i)} \tilde{X}_{jk}^{(i)} + \bm{\beta}_z^{(i)} \bm{Z}_{jk} 
\end{align*}

\end{enumerate}

\noindent Repeat the above steps $I$ times to obtain $I$ estimates of $\hat{\bm{\beta}}^{(i)}$ and $\widehat{Var}(\hat{\bm{\beta}}^{(i)})$. We obtain a final estimate of the variance of $\hat{\bm{\beta}}$ by taking an average of the naive variance and empirical variance from the pseudo datasets such that

\begin{align}
\label{var}
\widehat{Var}(\hat{\bm{\beta}}) & = \sum_{i=1}^I \frac{\widehat{Var}(\hat{\bm{\beta}}^{(i)})}{I} + \sum_{i=1}^I \frac{(\hat{\bm{\beta}}^{(i)}-\bar{\bm{\beta}}^{(i)})^T(\hat{\bm{\beta}}^{(i)}-\bar{\bm{\beta}}^{(i)})}{I-1},
\end{align}

\noindent where $\bar{\bm{\beta}}^{(i)} = \sum_{i=1}^I \frac{\hat{\bm{\beta}}^{(i)}}{I}$. This variance estimator takes into account the variation due to estimating the first moment random and fixed effects parameters, $\bm{\theta}$ and $\bm{\xi}$, and ignores the variation due to estimating the second moment parameters in $\bm{V}$. Our simulation studies in Section \ref{secsimstudy} show robust performance of this proposed variance estimator. An alternative variance estimator used by Boshuizen et al. (2007) inflates the second term in equation (\ref{var}) by $(I+1)/I$. Operating characteristics of this estimator are explored in Section F of the Supporting information. Relative to the estimator in equation (\ref{var}), this inflated estimator provided more accurate variance estimators for large ORs but frequently overestimated the variance for weak and moderate ORs (i.e. 0.67 to 1.5).  Since weak and moderate ORs are more common in epidemiologic studies, we used equation (\ref{var}) for simulations in Section 3.

\section{Simulation study}
\label{secsimstudy}

For the base setting, we generated data for $m=5$ contributing studies with a single covariate $\bm{Z}_{jk}=W_{jk} \sim N(0,1)$. We assumed each study contained 500 subjects, 50 of whom had a second biomarker measurement at the central laboratory. We then generated $X_{jk}$ values per model (\ref{defineX}), setting the study-specific intercepts $\bm{\alpha}_0=(-0.1, -0.2, -0.3, -0.4, -0.5)$, the single covariate $\tau=1$, and true measurement variability $\sigma^2_x=5$. We then generated the lab-specific measurements per model (\ref{defineH}), setting the lab-specific measurement variability values $(\sigma^2_0,\sigma^2_1,\sigma^2_2,\sigma^2_3,\sigma^2_4,\sigma^2_5)=(2,3,4,5,6,7)$ and generating lab-specific random effects by $\xi_d \sim N(0, \sigma^2_{\xi}=3)$. In model (\ref{logit}), we set $\beta_z=\log(1.25)$ and generated study-specific values of $\beta_{0j}$ that corresponded to a desired average prevalence level of 10\% across all simulations. At each OR considered, including (0.5, 0.67, 0.8, 1.25, 1.5, 2.0), we completed 1000 simulations. We used $I=20$ pseudo datasets to estimate $\widehat{Var}(\hat{\beta}_x)$.

We compared our method, which treats the local and central laboratory biomarker measurements as repeated measures and thus is referred to as the repeated measures method, to three other procedures to provide context for its performance. A `naive' procedure obtains $\beta_x$ estimates by fitting model (\ref{logit}) with local laboratory values $H_{jjk}$ in place of $X_{jk}$. The `true values' procedure assumes all $X_{jk}$ values are available and directly fits model (\ref{logit}). The `known parameters' procedure fits model (\ref{tildelogit}) but assumes $(\bm{\theta},\bm{\xi},\bm{\sigma}^2)$ are known when computing the calibrated measurements $\tilde{X}_{jk}$. The known parameters procedure provides context for how much additional variance or point estimate bias is incurred by estimating $\hat{\bm{\theta}}$, $\hat{\bm{\xi}}$, and $\hat{\bm{\sigma}}^2$ when compared to the repeated measures approach. Note that only the naive and our repeated measures approaches can be fit in practice with available data. All four methods were compared with regard to mean percent bias in $\hat{\beta}_x$ over the simulation replicates, mean squared error (MSE), and coverage rate, defined as the percentage of replicates whose 95\% confidence intervals covered the true $\beta_x$. We denote the $\hat{\beta}_x$ estimate resulting from the naive, true values, known parameters, and repeated measures approach as $\hat{\beta}_N$, $\hat{\beta}_T$, $\hat{\beta}_K$, and $\hat{\beta}_R$, respectively. 

We first compared the procedures over varying disease prevalence, including 5\%, 10\%, and 25\% (Table 1). For each prevalence, the naive method performed poorly, with the percent bias of $\hat{\beta}_x$ ranging between -51\% and -60\%. As expected, the true values approach yielded approximately unbiased point estimates and confidence interval coverage rates close to the nominal 95\%.  The known parameters and our repeated measures approaches provided similar point estimates. The percent bias for $\hat{\beta}_R$ was always different from the percent bias of $\hat{\beta}_K$ by a margin of 0.3\% to 2.1\%. Neither method was uniformly less biased over the range of ORs considered. The standard errors (SEs) of $\hat{\beta}_K$ were smaller than the SEs of $\hat{\bm{\beta}}_R$, and the coverage rates of $\hat{\beta}_K$ tended to be slightly closer to the nominal 95\% than those of $\hat{\beta}_R$. As disease prevalence increased, $\hat{\beta}_K$ and $\hat{\beta}_R$ experienced increasing downward bias for large effects. For instance, at $\beta_x=\log(2.0)$, the percent bias of ($\hat{\beta}_K$,$\hat{\beta}_R$) decreased from (-3.7\%,-2.4\%) to (-7.6\%,-6.9\%) as the prevalence increased from 5\% to 25\%. The coverage rates for $\hat{\beta}_R$ also experienced slight depression for large effects. As the prevalence increased from 5\% to 25\% for $\beta_x=\log(2.0)$, the $\hat{\beta}_R$ coverage rate decreased from 0.95 to 0.91. When $\beta_x$ is within $\log(0.67)$ to $\log(1.5)$, which corresponds to weak to moderate effect, the 95\% coverage rate from our repeated measures method  ranged from 0.95 to 0.97.

Table 1 also illustrates that the bias of the proposed method is not always in the same direction. Under the simulation parameters used, the bias in the $\beta_x$  estimate is positive given a weak association and negative given a strong association, which is not unexpected because the bias introduced from the approximation in (4) is non-linear, as shown in Section A of the Supporting information.

Under the base settings defined at the beginning of this section, we also plotted average percent bias in $(\hat{\bm{\theta}},\hat{\bm{\sigma}}^2)$ and the average difference $ \bm{\xi}-\hat{\bm{\xi}}$ across the simulation replicates (Supporting information Section G). The absolute percent bias among the fixed effects was consistently less than 3\%. The bias in the study-specific intercepts $\hat{\bm{\alpha}}_0$ was slightly larger than the bias of $\hat{\tau}$, but this is anticipated given that all data contributes to the estimation of $\hat{\tau}$ while only the appropriate study-specific data contributes to the estimation of $\hat{\bm{\alpha}}_0$. The percent bias of each component of $\hat{\bm{\sigma}}^2$ was also close to zero on average and consistently less than 3\%. Similarly, the mean difference in $\bm{\xi} -\hat{\bm{\xi}}$ was nearly zero across the simulation replicates, indicating the the random effects were well estimated. Given that $(\hat{\bm{\theta}}, \hat{\bm{\xi}}, \hat{\bm{\sigma}}^2 )$ were well-estimated, it is not surprising that $\hat{\beta}_K$ and $\hat{\beta}_R$ were similar on average.

We also varied the percentage of subjects in the calibration subset, considering levels of 5\%, 8\%, and 15\% (Figure 1). As the percentage of individuals included in the calibration subset increased, $\hat{\beta}_K$ and $\hat{\beta}_R$ and their MSEs became more similar on average.

The impact of increasing the number of studies within $m=(2,4,8)$ was tested (Figure 2). As $m$ increased, $\hat{\beta}_K$ and $\hat{\beta}_R$ performed increasingly similar, due in part to the increasing overall sample size improving the $(\hat{\bm{\theta}}, \hat{\bm{\xi}}, \hat{\bm{\sigma}}^2 )$ estimates. For $\beta_x=(\log(0.5),\log(2.0))$ for $m=2$ studies, the MSE of $\hat{\beta}_R$ was nearly four times larger than the MSE of $\hat{\beta}_K$. These MSEs were increasingly similar as the number of studies increased to $m=4$ or $m=8$. 

We also investigated the method when the normality assumption was violated. If the distribution of the biomarker data is fat-tailed or skewed, $\epsilon_{xjk}$ will not be normally distributed. Two non-normal distributions for $\epsilon_{xjk}$ were tested: the uniform distribution ($-\sqrt{5},\sqrt{5}$) and the skew normal distribution (shape parameter $\alpha=1$). These distributions were parameterized to have the same mean and variance as $\epsilon_{xjk}$ in the base simulation. All other simulation parameters remained the same. Figure S5 of the Supporting information compares the simulation results from normal, uniform, and skew normal error terms. We observe that the percent bias and MSE in the $\beta_x$ estimates for normal errors are quite similar to those from the non-normal errors.

We also investigated whether a nonzero cluster mean impacts the method's performance, i.e., $\xi_d \sim N(a,\sigma^2_{\xi})$ where $a \neq 0$. As described in Section G of the Supporting information, $a$ is absorbed by the intercept in model (2) and does not impact the estimators. Simulations support this conclusion (see Figure S6 of the Supporting information).

We analyzed how changing the number of pseudo datasets impacted the resulting coverage rates for the repeated measures approach for $I=(5,10,20,50)$ (Table 2). For $\beta_x$ near the null, we observed similar coverage rates regardless of the value of $I$. At the most extreme ORs, increasing $I$ slightly improved the coverage rate. For example, at $\beta_x=\log(0.5)$, increasing $I$ from 5 to 50 improved the coverage rate from 90.2\% to 92.5\%. 

The repeated measures approach was compared against the one and two stage approaches of Sloan et al (2018), with results shown in Supporting information, Table S6. Compared to the repeated measures estimates, the one and two stage estimates have larger magnitude of percent bias, smaller SEs, smaller MSEs, and less accurate 95\% coverage rates compared with the repeated measures estimates. For instance, for OR=1.5, the percent bias, SE, MSE, and 95\% CI coverage rate of the two stage method are (-13.9\%, 0.069, 0.0062, 0.81) respectively while the corresponding measures for the repeated measures approach are (5.0\%, 0.086, 0.0081, 0.95). 

\begin{figure}[h]
\centering
\includegraphics[width=15cm]{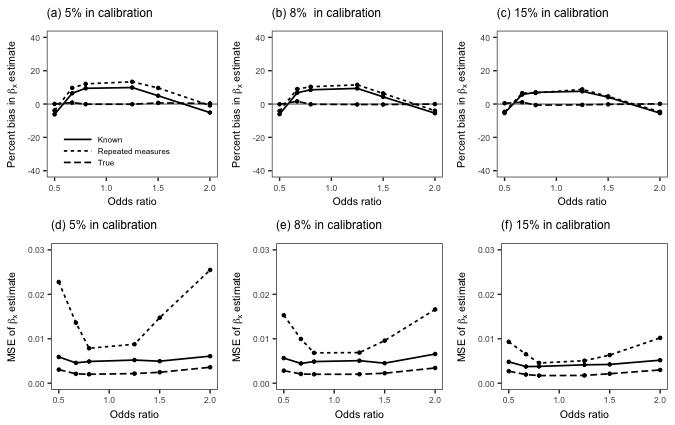}
\caption{Comparison of approaches as the percentage of subjects included in the calibration subset increases. Panels (a) to (c) illustrate the percent bias in the $\beta_x$ estimate while panels (d) to (f) depict their MSE. The x axis gives the exponentiated $\beta_x$ terms. The `true values' procedure assumes all $X_{jk}$ values are available. The `known parameters' procedure fits model (\ref{tildelogit}) but assumes $(\bm{\theta},\bm{\xi},\bm{\sigma}^2)$ are known when computing the calibrated measurements $\tilde{X}_{jk}$. The repeated measures approach estimates all necessary parameters. Five contributing studies are assumed, each with 500 subjects. Operating characteristics for the naive approach are provided in Table S2 of the Supporting information.}
\label{figvaryncal}
\end{figure}

\begin{figure}[h]
\centering
\includegraphics[width=15cm]{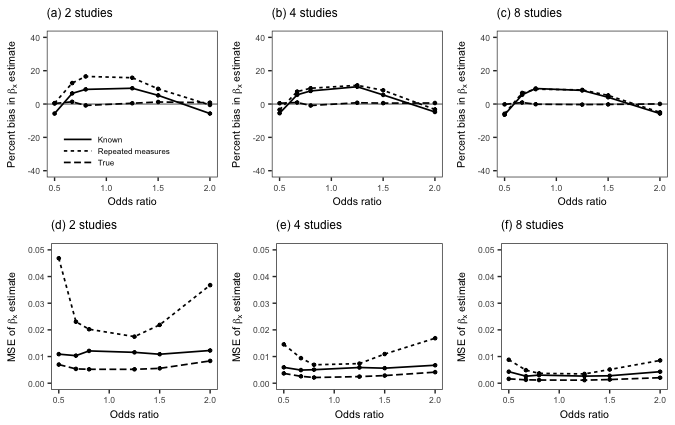}
\caption{Comparison of approaches as the number of studies in the pooled analysis increases. Panels (a) to (c) illustrate the percent bias in the $\beta_x$ estimate while panels (d) to (f) depict their MSE. The x axis gives the exponentiated $\beta_x$ terms. The `true values' procedure assumes all $X_{jk}$ values are available. The `known parameters' procedure fits model (\ref{tildelogit}) but assumes $(\bm{\theta},\bm{\xi},\bm{\sigma}^2)$ are known when computing the calibrated measurements $\tilde{X}_{jk}$. The repeated measures approach estimates all necessary parameters. Five contributing studies are assumed, each with 500 subjects and 50 subjects in the calibration subset. Operating characteristics for the naive approach are provided in Table S3 of the Supporting information.}
\label{figvarynstud}
\end{figure}

\begin{sidewaystable}[p]
\small
\caption{Comparison of operating characteristics under varying disease prevalence for naive approach ($\hat{\beta}_N$),  `true' approach  ($\hat{\beta}_T$), `known parameters' approach ($\hat{\beta}_K$), and the proposed repeated measures approach ($\hat{\beta}_R$).*}
\centering
\begin{tabular}{cccccccccccccccc}
\hline
           &           & \multicolumn{4}{c}{Percent Bias (Standard Error) of $\hat{\beta}$}          &  & \multicolumn{4}{c}{MSE}           &  & \multicolumn{4}{c}{95\% Coverage Rate} \\ \cline{3-6} \cline{8-11} \cline{13-16} 
Prevalence & $\beta_x$     & $\hat{\beta}_N$            & $\hat{\beta}_{T}$            & $\hat{\beta}_{K}$           & $\hat{\beta}_{R}$    &       & $\hat{\beta}_N$      & $\hat{\beta}_{T}$    & $\hat{\beta}_{K}$    & $\hat{\beta}_{R}$ &   & $\hat{\beta}_N$       & $\hat{\beta}_{T}$     & $\hat{\beta}_{K}$     & $\hat{\beta}_{R}$      \\ \hline
5          & log(0.50) & -57.4 (0.034) & 0.7 (0.060)  & -3.6 (0.072) & -2.3 (0.114) &  & 0.1597 & 0.0037 & 0.0058 & 0.0133 &  & 0.00   & 0.95   & 0.94   & 0.95   \\
           & log(0.67) & -52.3 (0.037) & 0.5 (0.054)  & 7.5 (0.079)  & 9.1 (0.104)  &  & 0.0463 & 0.0029 & 0.0071 & 0.0121 &  & 0.00   & 0.95   & 0.95   & 0.96   \\
           & log(0.80) & -51.2 (0.040) & 0.2 (0.057)  & 9.6 (0.087)  & 11.7 (0.096) &  & 0.0146 & 0.0032 & 0.0081 & 0.0099 &  & 0.16   & 0.95   & 0.95   & 0.96   \\
           & log(1.25) & -51.6 (0.039) & 0.3 (0.057)  & 8.4 (0.084)  & 10.3 (0.093) &  & 0.0148 & 0.0033 & 0.0075 & 0.0092 &  & 0.17   & 0.96   & 0.95   & 0.96   \\
           & log(1.50) & -52.8 (0.037) & 0.0 (0.058)  & 6.3 (0.081)  & 8.2 (0.102)  &  & 0.0473 & 0.0034 & 0.0073 & 0.0115 &  & 0.00   & 0.95   & 0.96   & 0.96   \\
           & log(2.00) & -57.5 (0.036) & 0.1 (0.063)  & -3.7 (0.076) & -2.4 (0.117) &  & 0.1602 & 0.0040 & 0.0065 & 0.0140 &  & 0.00   & 0.95   & 0.94   & 0.95   \\
           & $\beta_x$     & $\hat{\beta}_N$            & $\hat{\beta}_{T}$            & $\hat{\beta}_{K}$           & $\hat{\beta}_{R}$    &       & $\hat{\beta}_N$      & $\hat{\beta}_{T}$    & $\hat{\beta}_{K}$    & $\hat{\beta}_{R}$ &   & $\hat{\beta}_N$       & $\hat{\beta}_{T}$     & $\hat{\beta}_{K}$     & $\hat{\beta}_{R}$     \\
10         & log(0.50) & -58.8 (0.028) & 0.0 (0.051)  & -6.4 (0.059) & -5.0 (0.111) &  & 0.1671 & 0.0026 & 0.0055 & 0.0135 &  & 0.00   & 0.96   & 0.94   & 0.93   \\
           & log(0.67) & -53.7 (0.029) & 0.0 (0.045)  & 4.2 (0.062)  & 5.3 (0.085)  &  & 0.0482 & 0.0020 & 0.0041 & 0.0077 &  & 0.00   & 0.96   & 0.95   & 0.96   \\
           & log(0.80) & -51.0 (0.030) & 1.0 (0.043)  & 9.7 (0.065)  & 11.2 (0.074) &  & 0.0138 & 0.0019 & 0.0046 & 0.0061 &  & 0.04   & 0.95   & 0.95   & 0.97   \\
           & log(1.25) & -51.4 (0.030) & 0.0 (0.042)  & 9.3 (0.065)  & 10.7 (0.073) &  & 0.0141 & 0.0018 & 0.0046 & 0.0060 &  & 0.05   & 0.96   & 0.96   & 0.97   \\
           & log(1.50) & -53.5 (0.030) & -0.3 (0.047) & 4.7 (0.063)  & 5.5 (0.087)  &  & 0.0480 & 0.0022 & 0.0043 & 0.0081 &  & 0.00   & 0.95   & 0.95   & 0.96   \\
           & log(2.00) & -58.2 (0.030) & 0.5 (0.058)  & -5.0 (0.065) & -4.1 (0.111) &  & 0.1637 & 0.0034 & 0.0055 & 0.0130 &  & 0.00   & 0.94   & 0.94   & 0.94   \\
           & $\beta_x$     & $\hat{\beta}_N$            & $\hat{\beta}_{T}$            & $\hat{\beta}_{K}$           & $\hat{\beta}_{R}$    &       & $\hat{\beta}_N$      & $\hat{\beta}_{T}$    & $\hat{\beta}_{K}$    & $\hat{\beta}_{R}$ &   & $\hat{\beta}_N$       & $\hat{\beta}_{T}$     & $\hat{\beta}_{K}$     & $\hat{\beta}_{R}$     \\
25         & log(0.50) & -59.9 (0.025) & -0.3 (0.045) & -8.3 (0.052) & -7.7 (0.102) &  & 0.1729 & 0.0021 & 0.0061 & 0.0133 &  & 0.00   & 0.96   & 0.93   & 0.91   \\
           & log(0.67) & -54.8 (0.022) & 0.0 (0.037)  & 2.5 (0.047)  & 2.8 (0.072)  &  & 0.0498 & 0.0014 & 0.0023 & 0.0053 &  & 0.00   & 0.94   & 0.95   & 0.95   \\
           & log(0.80) & -51.9 (0.021) & 0.4 (0.031)  & 8.2 (0.045)  & 10.2 (0.057) &  & 0.0138 & 0.0010 & 0.0023 & 0.0037 &  & 0.00   & 0.95   & 0.94   & 0.97   \\
           & log(1.25) & -52.4 (0.022) & -0.6 (0.033) & 7.0 (0.048)  & 7.8 (0.059)  &  & 0.0142 & 0.0011 & 0.0026 & 0.0038 &  & 0.00   & 0.95   & 0.93   & 0.97   \\
           & log(1.50) & -54.4 (0.023) & 0.0 (0.038)  & 3.0 (0.049)  & 3.9 (0.074)  &  & 0.0492 & 0.0014 & 0.0026 & 0.0057 &  & 0.00   & 0.95   & 0.95   & 0.96   \\
           & log(2.00) & -59.5 (0.027) & 0.1 (0.049)  & -7.6 (0.057) & -6.9 (0.106) &  & 0.1706 & 0.0025 & 0.0060 & 0.0135 &  & 0.00   & 0.94   & 0.93   & 0.91   \\ \hline
\end{tabular}
\vspace{8mm} \begin{flushleft} *The naive procedure obtains $\beta_x$ estimates by fitting model (\ref{logit}) with local laboratory values. The `true values' procedure assumes all $X_{jk}$ values are available and directly fits model (\ref{logit}). The `known parameters' procedure fits model (\ref{tildelogit}) but assumes $(\bm{\theta},\bm{\xi},\bm{\sigma}^2)$ are known when computing the calibrated measurements $\tilde{X}_{jk}$. The repeated measures approach estimates all necessary parameters. Simulations assume five studies contribute to the analysis. Each study has 500 subjects, 50 of whom have a repeated central laboratory measurement. Standard error is the square root of the empirical variance over all replicates. \end{flushleft}
\label{tabvaryprev}
\end{sidewaystable}

\begin{table}[h]
\centering
\caption{95\% CI coverage rates for $\beta_x$ over 1000 simulations under $I$ total pseudo datasets. The simulation parameters are 10\% disease prevalence, 5 studies, 500 subjects in each study, and 50 subjects in each calibration subset.}
\begin{tabular}{ccccccc}
\hline
Number of       & \multicolumn{6}{c}{Odds ratio}          \\
pseudo datasets & 0.50 & 0.67 & 0.80 & 1.25 & 1.50 & 2.00 \\ \hline
I=5            & 90.2 & 95.1 & 96.4 & 96.7 & 94.6 & 90.9 \\
I=10            & 91.2 & 93.9 & 96.5 & 96.5 & 94.1 & 91.6 \\
I=20            & 91.4 & 95.0 & 96.0 & 96.6 & 96.4 & 91.3 \\
I=50            & 92.5 & 95.3 & 95.9 & 96.4 & 95.7 & 92.7 \\ \hline
\end{tabular}
\label{tabvaryI}
\end{table}

%%%%%%%%%%%%%%%%%%%%%%%%%%%%%%%%%%%%%%%%%%%%%%%%%%%%%%%%%%%%%%%%%%%%%%%
\section{Applied Example}

We examined the association between circulating 25(OH)D and risk of stroke using data pooled from three prospective cohort studies in the United States, including the Health Professionals Follow-Up Study (HPFS) [\cite{Wu11}], Nurses' Health Study I (NHS1) [\cite{Eliassen16}], and Nurses' Health Study II (NHS2) [\cite{Eliassen11}]. Additional information on each study is available elsewhere, but we briefly describe the demographic characteristics of each study [\cite{Grobbee90, Colditz97, Rich02}]. In 1986, the HPFS enrolled 51,529 male health professionals aged 40 to 75 years at baseline. The NHS1 began in 1976, enrolling 121,701 female nurses aged 30 to 55 years at baseline. Its younger counterpart, the NHS2, enrolled 116,671 female nurses in 1989 aged 25 to 42 years at baseline. Participants in each study completed biannual questionnaires providing information about diet, medical history, and lifestyle factors.

Each study gathered blood samples from a small subset of subjects and completed assays for 25(OH)D, among other biomarkers, between 1989 and 1997 (see Table S4 of the Supporting information). Participants with a prior cancer diagnosis were excluded from random selection. We excluded subjects from the pooled analysis if they were missing 25(OH)D measurements or stroke outcome data. In total, 3,768 subjects were included in the pooled analysis, which we regard as a random sample arising from the HPFS, NHS1, and NHS2 cohorts. We classified participants as stroke cases if they experienced nonfatal or fatal suspected or confirmed stroke event during study follow-up. While this data could be approached from a time-to-event perspective, we have opted to transform stroke into a binary outcome and control for follow-up time to enable comparison with previous pooled analyses for this data. The cumulative incidence of stroke in the final pooled cohort was 4.8\%, with individual cumulative incidences of 5.8\%, 9.6\%, and 1.5\% in the separate HPFS, NHS1, and NHS2 cohorts. 

To address inter-study assay variability, each cohort randomly re-assayed 29 blood samples from 2011 to 2013 at a central laboratory (Heartland Assays LLC, Ames, IA). Previous pooled analyses by Sloan et al. (2018, 2019) used aggregated and two-stage approaches to analyze the relationship between circulating 25(OH)D and stroke. These analyses adjusted for age at blood draw, years of follow-up after blood draw, smoking (never/ever), family history of myocardial infarction (yes/no), personal history of hypertension (yes/no), personal history of diabetes (yes/no), and BMI (greater than or less than or equal to 25 kg/m$^2$) [\cite{McCullough18}]. BMI was dichotomized using the same cutoff point as Sloan et al. (2018, 2019) to facilitate comparison of results.

We used the repeated measures approach introduced in this paper on the same data set, adjusting for the same covariates and analyzing the same association between 25(OH)D and stroke, and compared results (Table 3). Variance estimates from the resampling approach used $I=50$ pseudo datasets.

Our repeated measures approach suggests that a 20 nmol/L increase in circulating 25(OH)D corresponds to a 5.7\% reduction in stroke risk (OR 0.943; CI (0.780, 1.141)). This OR estimate is similar to those estimated under the internalized, full calibration, or two-stage method, which were estimated to be 0.951, 0.951, and 0.947 respectively. This could be due to the relatively large intra-lab correlation coefficients (ICC) in this example, defined as $\widehat{ICC}_d=\hat{\sigma}^2_d / (\hat{\sigma}^2_{\xi}+ \hat{\sigma}^2_d )$ and estimated to be $(\widehat{ICC}_0,\widehat{ICC}_1,\widehat{ICC}_2,\widehat{ICC}_3)=(0.85, 0.87, 0.89, 0.88)$. The difference between the repeated measures approach and the other approaches could be larger for smaller ICCs, especially if the ICC for the central laboratory is as small since the internalized, full calibration, and two-stage methods do not account for measurement error in the central laboratory measurement. 

The estimated SE of the coefficient estimate was slightly lower under our repeated measures approach compared to the other adjustment methods of Sloan et al. (2018). All estimates from calibration methods were moderately different from the naive estimate, which reported an OR estimate biased toward the null of 0.98.

\begin{table}[h]
\centering
\label{tabcompareOR}
\caption{Point and interval estimates of OR for 25(OH)D effect on stroke from the naive and repeated measures methods obtained by Sloan et al. (2018), comparing with those using the internalized, full calibration and two-stage method.*OR estimates correspond to a 20 nmol/L increase in circulating 25(OH)D. The linear and logistic regression models in (1) and (3) are adjusted for BMI, smoking history, family history of myocardial infarction, age at blood draw, diabetes diagnosis, follow-up duration after blood draw, and high blood pressure diagnosis. }
\begin{tabular}{lcc}
\hline
Method            & $\beta_x$ (SE)*     & OR (95\% CI)           \\ \hline
Naive             & -0.020 (0.077) & 0.980 (0.842, 1.140) \\
Internalized      & -0.050 (0.100) & 0.951 (0.782, 1.157) \\
Full calibration  & -0.050 (0.101) & 0.951 (0.780, 1.159) \\
Two-stage         & -0.054 (0.102) & 0.947 (0.776, 1.157) \\
Repeated measures & -0.058 (0.097) & 0.943 (0.780, 1.141) \\ \hline
\end{tabular}
\end{table}

\section{Discussion}

In this paper, we proposed a repeated measures approach to estimate the OR associated with a binary disease outcome and continuous biomarker exposure covariate for data pooled across multiple studies. We assumed that the biomarker displayed inter-study measurement heterogeneity related to laboratory and/or assay variability, thus precluding usage of a direct participant level meta-analysis. The repeated measures framework, implemented here as a LMM, allows us to account for correlation between (1) measurements repeated at the same laboratory on different individuals and (2) measurements repeated on the same individual at different labs, in estimation of the biomarker value. The inclusion of study-specific intercepts and covariates in model (2) and laboratory-specific measurement error variances enable the method to flexibly address measurement differences.

Our simulation studies demonstrate that $\hat{\beta}_x$ estimates from the repeated measures approach had satisfactory operating characteristics that greatly improve upon those from a naive approach. We also compared the repeated measures approach to the known parameters approach, where we assumed the fixed effects, random effects, and variance terms were known. These two methods provided similar point estimates, indicating the fixed effects, random effects, and variance terms were well-estimated by the LMM.

Our simulation work also demonstrated that our method's performance was most sensitive to the number of subjects included in the calibration subset, with increasing calibration subset sizes improving estimation. We recommend enriching the calibration subset to the largest extent allowed by study resources.

Previous methods for pooled biomarker data with inter-study variability by Sloan et al. (2018) regarded the central laboratory measurements as a gold standard. The study-specific calibration models were implemented as Berkson measurement error models across different data subsets. However, most central laboratories are chosen for logistical convenience rather than their status as a gold standard. In contrast, the proposed repeated measures method takes into account measurement error in both the central and local laboratory measurements. In general, the repeated measures approach has relatively few drawbacks. It is somewhat computationally intensive because computing $\hat{\bm{V}}^{-1}$ requires the inversion of a matrix whose dimensions are on the order of the number of biomarker measurements. However, because most of the entries of $\hat{\bm{V}}$ are zero, this inversion is handled well by most statistical programs. Computational speed can also be improved by a Cholesky decomposition. This approach works best when the number of clusters (laboratories) is at least four; having fewer clusters impedes estimation of $\hat{\sigma}^2_{\xi}$ and the random effects. 

We treated stroke as a binary outcome in the applied example while controlling for follow-up time. A future direction could involve extending this method to the time-to-event setting.

\backmatter

%  This section is optional.  Here is where you will want to cite
%  grants, people who helped with the paper, etc.  But keep it short!

\section*{Acknowledgements}
We thank Mitchell Gail and the reviewers for their insightful comments. We also thank the Circulating Biomarkers and Breast and Colorectal Cancer Consortium team (R01CA152071, PI: Stephanie Smith-Warner; Intramural Research Program, Division of Cancer Epidemiology and Genetics, National Cancer Institute: Regina Ziegler) for conducting the calibration study in the vitamin D examples.  A.S. was supported by NIH grant T32-NS048005 and M. W. was supported by NIH/NCI grant R03CA212799. 

%  Not included in the original version of this paper!

%  Here, we create the bibliographic entries manually, following the
%  journal style.  If you use this method or use natbib, PLEASE PAY
%  CAREFUL ATTENTION TO THE BIBLIOGRAPHIC STYLE IN A RECENT ISSUE OF
%  THE JOURNAL AND FOLLOW IT!  Failure to follow stylistic conventions
%  just lengthens the time spend copyediting your paper and hence its
%  position in the publication queue should it be accepted.

%  We greatly prefer that you incorporate the references for your
%  article into the body of the article as we have done here 
%  (you can use natbib or not as you choose) than use BiBTeX,
%  so that your article is self-contained in one file.
%  If you do use BiBTeX, please use the .bst file that comes with 
%  the distribution.

\section*{Data Availability Statement}
Research data are not shared.

\section*{Supporting Information}
Web Appendices, Tables, and Figures referenced in Sections 2.2, 2.3, 2.4, 2.5, 2.6, and 3 are available with this paper at the Biometrics website on Wiley Online Library.
\vspace*{-8pt}

\label{lastpage}


\begin{thebibliography}{}

%\bibitem[\protect\citeauthoryear{Akash and Tirky}{1988}]{b24} 
%Akash, F. J. and Tirky, T. (1988). Proper multivariate conditional
%autoregressive models for spatial data analysis. {\it Biometrics} {\bf 196,} 173.

\bibitem[\protect\citeauthoryear{Barake et al.}{2012}]{Barake12} Barake, M., Daher, R. T., Salti, I., Cortas, N. K., Al-Shaar, R. Habib, H., et al (2012). 25-hydroxyvitamin D assay variations and impact on clinical decision making. {\it The Journal of Clinical Endocrinology \& Metabolism} {\bf 97(3), } 835-843.

\bibitem[\protect\citeauthoryear{Boshuizen et al.}{2007}]{Boshuizen07} Boshuizen, H. C., Lanti, M., Menotti, A., Moschandreas, J., Tolonen, H., Nissinen, A., Nedeljkovic, S., Kafatos, A., Kromhout, D., et al (2007). Effects of past and recent blood pressure and cholesterol level on coronary heart disease and stroke mortality, accounting for measurement error {\it American Journal of Epidemiology} {\bf 165(4), } 398-409.

\bibitem[\protect\citeauthoryear{Carroll et al.}{2006}]{Carroll06} Carroll, R., Ruppert, D., Stefanski, L., Crainiceanu, C. (2006). Measurement error in nonlinear models: a modern perspective; 2nd ed. {\it Chapman and Hall}.

\bibitem[\protect\citeauthoryear{Colditz et al.}{1997}]{Colditz97} Colditz, G. A., Manson, J. E., Hankinson, S. E (1997). The Nurses' Health Study: 20-year contribution to the understanding of health among women. {\it Journal of Women's Health} {\bf 6(1), } 49-62.


\bibitem[\protect\citeauthoryear{Crowe et al.}{2014}]{Crowe14} Crowe, F.L., Appleby, P.N., Travis, R.C., Barnett, M., Brasky, T.M., Bueno-de-Mesquita, H.B., et al. (2014). Circulating fatty acids and prostate cancer risk: individual participant meta-analysis of prospective studies. {\it Journal of the NCI} {\bf 106(9),} 1-10.

\bibitem[\protect\citeauthoryear{Debray et al.}{2013}]{Debray13} Debray, T. P. A., Moons, K. G. M, Abo-Zaid, G. M. A., Koffijberg, H., Riley, R.D. (2014). Individual participant data meta-analysis for a binary outcome: one-stage or two-stage? {\it PloS one} {\bf 8(4),} e60650.


\bibitem[\protect\citeauthoryear{Eliassen et al.}{2011}]{Eliassen11} Eliassen, A. H., Spiegelman, D., Hollis, B. W., Horst, R. L., Willett, W. C., Hankinson, S. E. (2011). Plasma 25-hydroxyvitamin D and risk of breast cancer in the Nurses' Health study II. {\it Breast cancer research} {\bf 13(3)}.


\bibitem[\protect\citeauthoryear{Eliassen et al.}{2016}]{Eliassen16} Eliassen, A. H., Warner, E. T., Rosner, B., Collins, L. C., Beck, A. H., Quintana, L. M., et al (2016). Plasma 25-hydroxyvitamin D and risk of breast cancer in women followed over 20 years (2016). {\it Cancer research} {\bf 76(18),} 5423-5430.

\bibitem[\protect\citeauthoryear{Freedman et al.}{2011}]{Freedman11} Freedman, L. S., Midthune, D., Carroll, R. J., Tasevska, N., Schatzkin, A., Mares, J., Tinker, L., Potischman, N., Kipnis, V. (2011). Plasma 25-hydroxyvitamin D and risk of breast cancer in women followed over 20 years (2016). {\it American Journal of Epidemiology} {\bf 174(11),} 1238-1245.

\bibitem[\protect\citeauthoryear{Gail et al.}{2016}]{Gail16} Gail, M.H., Wu, J., Wang, M., Yaun, S.S., Cook, N.R., Eliassen, A.H., et al (2016). Calibration and seasonal adjustment for matched case–control studies of vitamin D and cancer. {\it Statistics in Medicine} {\bf 35(13),} 2133-2148.

\bibitem[\protect\citeauthoryear{Grobbee et al.}{1990}]{Grobbee90} Grobbee, D. E., Rimm, E. B., Giovannucci, G., Colditz, G., Stampfer, M., Willett, W. (2016). Coffee, caffeine, and cardiovascular disease in men. {\it New England Journal of Medicine} {\bf 323(15),} 1026-1032.

\bibitem[\protect\citeauthoryear{Henderson}{1975}]{Henderson75} Henderson, C. R. (1975). Best linear unbiased estimation and prediction under a selection model. {\it Biometrics} {\bf 31(2),} 423-447.


\bibitem[\protect\citeauthoryear{Key et al.}{2010}]{Key10} Key, T. J., Appleby, P. N., Allen, N. E., Reeves, G. K. (2010). Pooling biomarker data from different studies of disease risk, with a focus on endogenous hormones. {\it Cancer Epidemiology and Prevention} {\bf 19(4),} 960-965.

\bibitem[\protect\citeauthoryear{Key et al.}{2015}]{Key15} Key, T. J., Appleby, P. N., Travis, R. C., Albanes, D., Alberg, A. J., Barricarte, A.,et al. (2015). Carotenoids, retinol, tocopherols, and prostate cancer risk: pooled analysis of 15 studies. {\it The American Journal of Clinical Nutrition} {\bf 102(5),} 1142-1157.

\bibitem[\protect\citeauthoryear{McCullough et al.}{2018}]{McCullough18} McCullough, M. L., Zoltick, E. S., Weinstein, S. J., Fedirko, V., Wang, M., Cook, N.R., et al. (2018). Circulating Vitamin D and Colorectal Cancer Risk: An International Pooling Project of 17 Cohorts. {\it Journal of the National Cancer Institute} {\bf 111(2),} 158-169.


\bibitem[\protect\citeauthoryear{Rao}{1972}]{Rao72} Rao, C.R. (1972). Estimation of variance and covariance components in linear models. {\it Journal of the American Statistical Association} {\bf 67(337),} 112-115.

\bibitem[\protect\citeauthoryear{Rich-Edwards et al.}{2002}]{Rich02} Rich-Edwards, J. W., Spiegelman, D., Garland, M., Hertzmark, E., Hunter, D. J., Colditz, G. A. et al. (2002). Physical activity, body mass index, and ovulatory disorder infertility. {\it Epidemiology} {\bf 13(2),} 184-190.

\bibitem[\protect\citeauthoryear{Rosner et al.}{1989}]{Rosner89} Rosner, B., Willett, W. C., Spiegelman, D. (1989). Correction of logistic regression relative risk estimates and confidence intervals for systematic within-person measurement error. {\it Statistics in Medicine} {\bf 8(9),} 1051-1069.

\bibitem[\protect\citeauthoryear{Rubin}{2004}]{Rubin04} Rubin, D. B. (2004). Multiple imputation for nonresponse in surveys, volume 81. {\it John Wiley \& Sons}.



\bibitem[\protect\citeauthoryear{Sloan et al.}{2018}]{Sloan18} Sloan, A., Song, Y., Gail, M.H., Ziegler, R.G., Smith-Warner, S.A., Wang, M. (2018). Design and analysis considerations for combining data from multiple biomarker studies. {\it Statistics in Medicine} {\bf 38(8),} 1303–1320.

\bibitem[\protect\citeauthoryear{Sloan et al.}{2019}]{Sloan19} Sloan, A., Smith-Warner, S.M., Ziegler, R. G., Wang, M. (2019). Statistical methods for biomarker data pooled from multiple nested case-control studies. {\it Biostatistics}  https://doi.org/10.1093/biostatistics/kxz051. [ePub ahead of print]


\bibitem[\protect\citeauthoryear{Smith-Warner et al.}{2006}]{Smith06} Smith-Warner, S. A., Spiegelman, D., Ritz, J., Albanes, D., Beeson, W. L., et al (2006). Methods for pooling results of epidemiologic studies: the pooling project of prospective studies of diet and cancer. {\it American Journal of Epidemiology} {\bf 163(11),} 1053–1064.


\bibitem[\protect\citeauthoryear{Tworoger et al.}{2006}]{Tworoger06} Tworoger, S.S., Hankinson, S.E. (2006). Use of biomarkers in epidemiologic studies: minimizing the influence of measurement error in the study design and analysis. {\it Cancer Causes \& Control} {\bf 17(7),} 889-899.

\bibitem[\protect\citeauthoryear{Wu et al.}{2011}]{Wu11} Wu, K., Feskanich, D., Fuchs, C. S., Chan, A. T., Willet, W. C., et al (2006). Interactions between palsa levels of 25-hydroxyvitamin D, insulin-like growth factor (IGF)-1 and C-peptide with risk of colorectal cancer. {\it PloS One} {\bf 6(12),} e28520.

\end{thebibliography}
\end{document}